\documentstyle[epsfig,natbib2,natbibmnfix]{mn2e}

\newcommand{\be}{\begin{equation}}
\newcommand{\ee}{\end{equation}}

\newcommand{\apj}{ApJ}

\newcommand{\mnras}{MNRAS}
\newcommand{\aap}{A\&A}

\newcommand{\apjl}{ApJL}

\newcommand{\nat}{Nature}

\def\ltsima{$\; \buildrel < \over \sim \;$}
\def\simlt{\lower.5ex\hbox{\ltsima}}
\def\gtsima{$\; \buildrel > \over \sim \;$}
\def\simgt{\lower.5ex\hbox{\gtsima}}
\def\sgra{Sgr~A$^*$}

\def\msun{{\,{\rm M}_\odot}}

\def\del#1{{}}

\title[]{Weighing the young stellar discs around \sgra}

\author[S.~Nayakshin, W. Dehnen, J. Cuadra \& R. Genzel] {\parbox{18cm}{Sergei
Nayakshin$^{1,2}$, Walter Dehnen$^1$, Jorge Cuadra$^{2}$ \& Reinhard
Genzel$^{3,4}$}\vspace{0.3cm}\\ $^1$Department of Physics \& Astronomy,
University of Leicester, Leicester, LE1 7RH, UK\\ $^2$Max-Planck-Institut
f\"{u}r Astrophysik, Karl-Schwarzschild-Stra\ss{}e 1, 85741 Garching bei
M\"{u}nchen, Germany\\ $^3$Max-Planck-Institut f\"{u}r Extraterrestrische
Physik (MPE), Garching bei M\"{u}nchen, Germany\\ $^4$Department of Physics,
University of California, Berkeley, CA 94720, USA }

\begin{document}

\maketitle

\begin{abstract}
It is believed that young massive stars orbiting \sgra\ in two stellar discs
on scales of $\sim 0.1-0.2$ parsecs were formed either farther out in the
Galaxy and then quickly migrated inward, or in situ in a massive
self-gravitating disc.  Comparing N-body evolution of stellar orbits with
observational constraints, we set upper limits on the masses of the two
stellar systems. These masses turn out to be few times lower than the expected
total stellar mass estimated from the observed young high-mass stellar
population and the standard galactic IMF.  If these stars were formed in situ,
in a massive self-gravitating disc, our results suggest that the formation of
low-mass stars was suppressed by a factor of at least a few, requiring a
top-heavy initial mass function (IMF) for stars formed near \sgra.
\end{abstract}

\renewcommand{\thefootnote}{\fnsymbol{footnote}}

\section{Introduction}
\label{intro}

\sgra\ is a $M_{\rm BH} \sim 3.5 \times 10^6 \msun$ super-massive black hole
(SMBH) in the centre of our Galaxy \citep[e.g., ][]{Schoedel02,Ghez03b}. Few
tens of close young massive stars dominate the energy output of the central
parsec of the Galaxy \citep{Krabbe95,Genzel03a}. The ages of the young stars
are estimated at $t = 6\pm 1$ million years \citep{Paumard05}. The origin of
these stars is an important mystery for astrophysics of Active Galactic Nuclei
(AGN). ``Normal'' modes of star formation at $0.1$ pc distances from a SMBH
are forbidden due to the huge tidal field of the central object. Star
formation in a massive gravitationally unstable accretion disc
\citep{Paczynski78,Kolykhalov80,Collin99,Goodman03} has been suggested
\citep{Levin03,Milosavljevic04,NC05}. Alternatively, the observed close young
stars could be remnants of a massive young star cluster whose orbit decayed
due to dynamical friction
\citep[e.g.,][]{Gerhard01,Kim03,Kim04,McMillan03,Gurkan05}.
  
A successful model for the origin of young stars in \sgra\ will have
to explain quantitatively not only the creation of the stars but also
their present day orbits. Almost all of the observed young stars in
\sgra\ belong to one of two {\em stellar} rings
\citep{Levin03,Genzel03a}.  Internal N-body disc evolution sets an
upper limit on the total mass of each of the stellar systems of the order 
of $M \simlt 3\times 10^5\msun$,
too high to be constraining for either of the models \citep{NC05}. In
addition to the internal disc thickening, discs precess in their mutual
(non axi-symmetric) potential and are warped with time
\citep{Nayakshin05}.  Stellar discs that end up too strongly warped or
thick will contradict the observations. \cite{NC05} suggested that
this sets an upper limit on the total stellar mass in the range of $M
\simlt (3-10)\times 10^4\msun$, but noted that more detailed tests are needed.

The goal of this paper is to perform such tests numerically.  Let
us estimate the magnitude of the warping effect. \cite{Nayakshin05}
calculated the rate at which a massless accretion disc is warped by a
massive ring inclined with respect to the (initially flat) disc at an
angle $\beta$. For radii $R$ much smaller or much larger than the ring
radius, $R_{\rm ring}$, the precession angular frequency $\omega_p(R)$
can be approximated as
\begin{equation}
\frac{\omega_p(R)}{\Omega_K(R)} \approx - \frac{3 M_{\rm ring}}{4
M_{\rm BH}} \;\cos \beta\; \frac{R^3 R_{\rm ring}^2} {\left[R^2 +
R_{\rm ring}^2\right]^{5/2}}\;.
\label{omegapa}
\end{equation}
Here $M_{\rm ring}$ is the ring mass, assumed to be much smaller than
the blackhole mass, $M_{\rm BH}$, and $\Omega_K(R)$ is the Kepler
circular frequency for the disc at radius $R$.  The period of circular
motion $3"$ ($1'' \approx 0.04\,$pc at the distance of \sgra) away
from \sgra, where most of the bright young stars are found, is around
2 thousand years.  Therefore, the stars at this location make about a
thousand revolutions in a few million years. Let $\Delta \gamma =
\omega_p t $ be the angle on which an annulus of the disc will precess
during this time: $\Delta \gamma \sim (M_{\rm ring}/M_{\rm BH}) \times
1000$. Very approximately, the disc warping will be noticeable when
$\Delta \gamma \sim 1$. Thus mass ratios $(M_{\rm ring}/M_{\rm BH})$
in excess of $10^{-3}$ or so may lead to warping of the stellar discs
in \sgra.

Equation \ref{omegapa} depends on the angle $\beta$, and would also
depend on the eccentricity of a stellar orbit if it were not assumed
to be zero in the derivation.  Stars are expected to be born on nearly
circular Keplerian orbits \citep[e.g.,][]{NC05}, i.e. with
eccentricity $e\sim H/R \ll 1$, for star formation in a
self-gravitating disc. On the other hand, \cite{Levin05} have recently
demonstrated that, due to a repetitive interaction with the IMBH,
stars  peeled off the IMBH-star cluster gain a substantial
eccentricity to their orbits. Even for the case of the circular IMBH
inspiral, stars acquire a mean eccentricity of $e\approx 0.5$.  We may
thus expect different mass limits for the two models of the origin of
the young stars near \sgra.

Reader not interested in the technical details of the simulations and
data comparison may simply look up the relevant limits and proceed to
the discussion of the results in \S \ref{sec:conclusions}.


\section{Numerical method}\label{sec:method}

To follow stellar orbits, we use the N-body package `NEMO'
\citep{Teuben95} with the orbit integrator `gyrfalcON'
\citep{Dehnen02}. The code calculates gravitational interaction of all
the particles and updates their velocities and positions. We model the
blackhole as a Plummer sphere with a core radius of $0.01"$. This
radius is much smaller than peri-centres of stellar orbits we
consider.  The stars are represented as particles with softening
radius of  $0.01''$ or less for some tests. Depending on the total mass of the
stars, we use between few hundred to few thousand particles for each
of the two stellar systems.

To set up the problem, we shall rely on the gross results of
\cite{Genzel03a}, and the more recent analysis of the data by \cite{Paumard05}.
For convenience only, we shall refer to the clock-wise rotating system
in the GC as a disc, and the counter clock-wise system as a ring. To
model the disc, we start with stars in a flat disc with the radial
extent from $R_{\rm in}=2''$ to $R_{\rm out}=5''$. The observed
counter clock-wise system contains fewer stars and it is harder to
assign the radial extent for this system \citep{Paumard05}. We used
two plausible guesses for the ring, therefore: one is a ring between
$R_{\rm in}=4''$ and $R_{\rm out}=5''$, and the other is a ring
between $R_{\rm in}=5''$ and $R_{\rm out}=7''$. The initial
inclination between the two systems is set at $\beta = 113^\circ$
\citep{Genzel03a,Paumard05}. Other parameters of initial conditions
specific to a model will be discussed below.

The simulations were ran until time $t\approx 3$ Million years. To
ensure numerical precision, the individual timesteps of the particles
were kept small enough, e.g. in the range of $0.06$ to $2$ years. The
angular momentum and total energy of the system were conserved to a
relative error of $10^{-3}$ or better. We did not include the
isotropic cluster of late type stars around \sgra\ in these
simulations because in the radial range of interest its mass is small
compared to that of \sgra, and it would not present any torques for
the discs because of the cluster's spherical symmetry.

At the end of a simulation, we fit the two stellar systems by a plane
in velocity space , $(v_x, v_y, v_z)$ . A plane is characterised by vector $\vec n$
orthogonal to the plane and normalised to unity, $\vec n^2 = 1$. The
best fitting planes are found via the reduced $\chi^2$ method for both
of the two stellar systems. The reduced $\chi^2$ for a fit is defined
as in \cite{Levin03}:
\begin{equation}
\chi^2 = \frac{1}{N_s-1}\; \sum_{i=1}^{N_{\rm s}} \frac{(v_x n_x + v_y n_y + v_z n_z)^2}{(\sigma_x n_x)^2 + (\sigma_y n_y)^2 + (\sigma_z n_z)^2}\;, 
\end{equation}
where $N_{\rm s}$ is the number of stars in the given stellar system,
$n_x, n_y$ and $n_z$ are the projections of $\vec n$, and $\sigma_x$,
$\sigma_y$ and $\sigma_z$ are the errors in stellar velocities in the
three directions. For simplicity we assume that these errors are
isotropic, i.e. each component is equal to $\sigma/\sqrt{3} \approx
40$ km/sec, where $\sigma = 70$ km/sec, the typical value for the absolute value of velocity error vector in \cite{Paumard05}.

Finally, to emulate the effects that the observational errors have on the $\chi^2$ fits through statistical scatter of the data around the mean, we added random Gaussian velocity kicks to each star's velocity vectors according to the errors $\sigma_x$, etc, at the end of the simulations. 

\section{Example runs for circular initial stellar orbits}\label{sec:circ}

As explained in \S \ref{sec:method}, we start with the
two stellar systems oriented as the observed best fitting planes
\citep{Paumard05}. The discs are populated by star particles with
surface density $\Sigma(R) \propto R^{-2}$ for $R_{\rm in} \le R \le
R_{\rm out}$. The initial velocity dispersion of stars with respect to
the local circular Kepler velocity is isotropic and small, $\langle
\vert\vec v- \vec v_K\vert\rangle = 0.017 v_K$, consistent with the
small initial gaseous disc thickness.

A convenient way to illustrate the results is via the `Aitoff map
projection' of the stellar orbits. Each stellar orbital plane can be
characterised by the inclination angle $i$ with respect to the
observer and the position angle $\phi$ which reflects the position of
the lines of the nodes of the orbit on the plane of the sky
\citep[e.g.][]{Schoedel02,Ghez03b,Eisenhauer05,Ghez05}.  
The normal vectors to the plane, introduced above, are related to these angles as $n_z = \cos i$, $n_x = \sin i \cos \phi$, and $n_y = \sin i \sin \phi$. The Aitoff's projection then shows the latitude, defined as $\pi/2 - i$, and longitude $=\phi$.

\begin{figure}
\centerline{\psfig{file=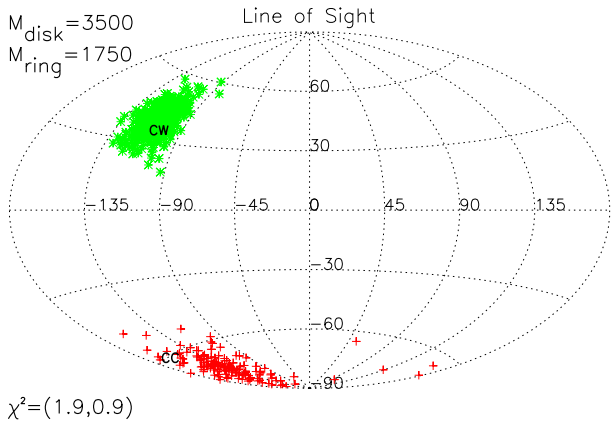,width=.48\textwidth,angle=0}}
\caption{The orientation angles of the stellar orbital planes, as described in the text. The
positions of the clockwise and counter-clockwise systems are shown
with symbols CW and CC, respectively. Total disc and ring
masses are labelled at the upper left corner. The values of the minimum
reduced $\chi^2$ fits to the (ring, disc) systems are displayed in the
lower left corner of the Figure. }
\label{fig:3la}
\end{figure}

Figure \ref{fig:3la} shows the state of the two stellar systems 3
million years into one of the simulations, e.g. the final state. The
total stellar masses in the disc and the ring systems (CW and CC
systems, respectively) used in this simulation are noted in the left
upper corner of the Figure.  The stars in the clock-wise system
(marked as CW on the Figure) are shown with green asterisks, whereas
the counter-clockwise system is shown with red crosses, marked as CC.
The lower left corner shows the resulting values of the minimum
reduced $\chi^2$ values as a (ring, disc) pair. These are comfortably
smaller than the observed values $\chi_{\rm cw}^2 = 2.5 $ and
$\chi_{\rm cc}^2 = 4.0 $ for the clock-wise and the counter clock-wise
stellar systems, respectively \citep{Paumard05}. Apparently such a
weak disc warping would not contradict observations.

\del{Figure \ref{fig:5la} shows the results of a calculation with same
initial conditions but with higher disc and ring masses. The
dispersion in the stellar angular momenta vectors increases, and the
corresponding $\chi^2$ values increase to $(\chi_2^2, \chi_1^2) = 3.6,
1.2)$. This still does not violate the observed values of  $(\chi_{\rm cw}^2, \chi_{\rm cc}^2) = (4.0, 2.5)$,
however. }


\begin{figure}
\centerline{\psfig{file=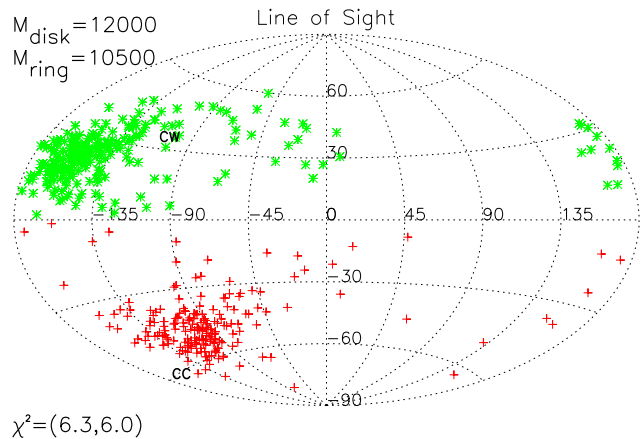,width=.5\textwidth,angle=0}}
\caption{Same as Figure \ref{fig:3la} but for much
higher disc and ring masses. Note that this stellar distribution is somewhat
inconsistent with the data, with the reduced $\chi^2$ values exceeding
the observed values of $\chi^2_{\rm cc, cw} = (4.0, 2.5)$.}
\label{fig:6la}
\end{figure}

Figure (\ref{fig:6la}), shows the same numerical experiment but with
the disc and ring masses both around $10^4\msun$.  As is clear from
the Figure, both systems suffer considerably from the gravitational
torques imposed on each other, and the resulting plane-parallel fits
to these are poorer than before. The reduced $\chi^2$ values are
larger than the observed values for both the clock-wise and the counter
clock-wise systems.

\begin{figure}
\centerline{\psfig{file=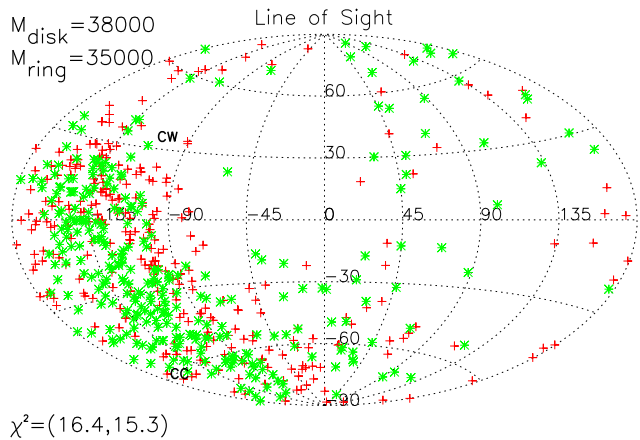,width=.5\textwidth,angle=0}}
\caption{Same as Figures \ref{fig:3la},\ref{fig:6la} but for the
largest values of the disc and ring masses considered. The reduced
$\chi^2$ values for both the ring and the disc are much larger than
the observed values. Such a stellar distribution cannot be classified
as consisting of two discs at all.}
\label{fig:7la1}
\end{figure}

Figure \ref{fig:7la1}  presents the Aitoff map for the most massive case that we have
explored, with the disc and ring masses both around $3.5\times
10^4\msun$. The resulting $\chi^2$ values
are quite large and are clearly inconsistent with the data. Note that
the degree of mixing occurring in this simulation is extreme, and some
of the particles that originally belonged to the clockwise disc are now
classified as members of the counter-clockwise system. A stellar
system with such poor values of $\chi^2$ may not even remotely be
considered as consisting of two or one stellar discs, obviously.

\begin{figure}
\centerline{\psfig{file=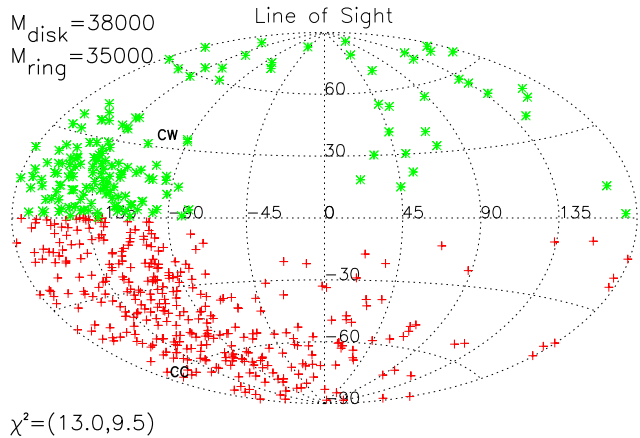,width=.5\textwidth,angle=0}}
\caption{Same test as shown in Figure \ref{fig:7la1}, but with stars
divided into populations based on whether they rotate clock-wise or
counter clock-wise. Note that the resulting $\chi^2$ values are
smaller but still much larger than the observed values.}
\label{fig:7la}
\end{figure}


In this latest example we have seen that stars from one system may
evolve on orbits more consistent with the other system, and it would
then be improper to continue to assign them to the disc of their birth
place. Certainly observationally such practice is impossible as it is
not a priory known in which system the stars originated. Instead,
one divides the stars on the clock-wise and the counter clock-wise
ones \citep{Genzel03a}.  To make a maximally fair comparison of the
simulations with the data, at the end of the simulation, we assign  particles to either the clock-wise or the counter clock-wise systems based on
the sign of $J_z = (x v_y - y v_x)$, where $x,y$ are the coordinates of the star.  The resulting change
is illustrated in Figure \ref{fig:7la}.  The Figure demonstrates that when the stellar systems are warped 
to the degree that their orbits diffuse into one another's phase space, even a more careful division of orbits still fails to produce systems as well defined as those observed to exist near \sgra.


\section{Results}\label{sec:results}

\subsection{Circular initial orbits}

Table \ref{tab:cl} lists results of several runs with the larger
stellar ring (i.e., $R=5"-7"$). The first two column show the total
stellar mass of the clock-wise (the disc) and the counter clock-wise
(the ring) systems in Solar masses. The next column shows an
identifier of the run (CL stands for "circular large"). Next two
columns list the best fitting $\chi^2$ values. The last column
contains a "+" if the $\chi^2$ values are smaller than the observed
ones, and a "-" in the opposite case. Clearly this is not to be taken
literally as in some cases the obtained $\chi^2$ are just slightly larger than the
observed ones.

\begin{table}
\caption{Simulations results for circular initial orbits with the larger "ring", with $R=5"-7"$. 
"fit quality" below, $+$ or $-$ sign, indicates whether the reduced $\chi^2$ values are smaller or larger than the observed values. Clearly this is just a rough measure of the model's feasibility.}

\begin{tabular}{@{}lllrrr@{}}
\hline
$M_{\rm disc}$ & $M_{\rm ring}$ & run & $\chi^2_{\rm disc}$ & $\chi^2_{\rm ring}$ & fit quality$^a$\\
\hline
3500 & 1750 & CL1 & 0.9 & 1.9 & +\\  
6300 & 1750 & CL2 & 1.6 & 1.0 & + \\  
12000 & 1750 & CL3 & 1.4 & 5.6 & - \\  
35000 & 1750 & CL4 & 3.2 & 12.7 & - \\ 
12000 & 3500 & CL5 & 1.3 & 3.7 & + \\ 
12000 & 10500 & CL6 & 6.0 & 6.3 & - \\ 
38000 & 10500 & CL7 & 6.2 & 11.5 & -\\ 
38000 & 35000 & CL8 & 9.5 & 13.0 & - \\ 
\hline
\label{tab:cl}
\end{tabular}
\end{table}

\begin{table}
\caption{Simulations results for circular initial orbits with the smaller ring, $R=4"-5"$.}
\begin{tabular}{@{}lllrrr@{}}
\hline
$M_{\rm disc}$ & $M_{\rm ring}$ & run & $\chi^2_{\rm disc}$ & $\chi^2_{\rm ring}$ & fit quality\\
\hline
3500 & 1750 & CS1 & 0.9 & 1.5 & +\\  
6300 & 1750 & CS2 & 1.2 & 1.0 & + \\  
12000 & 1750 & CS3 & 1. & 5.7 & - \\  
35000 & 1750 & CS4 & 3.6 & 17.6 & - \\ 
12000 & 3500 & CS5 & 1.2 & 4.6 & - \\ 
12000 & 10500 & CS6 & 4.2 & 7.0 & - \\ 
38000 & 10500 & CS7 & 8.7 & 16.0 & -\\ 
38000 & 35000 & CS8 & 10.0 & 10.8 & - \\ 
\hline
\label{tab:cs}
\end{tabular}
\end{table}

In a similar fashion, the results of the tests with a smaller stellar ring (the counter clock-wise system, $R = 4"-5"$) are presented in Table \ref{tab:cs}.  As one can see, the two tables are in general similar. 
The combined results of the small and larger scale ring tests are best described as these limits:
\begin{eqnarray}
M_{\rm disc} \simlt 1.5 \times 10^4 \msun\\
M_{\rm ring} \simlt 1. \times 10^4 \msun \;.
\label{circ}
\end{eqnarray}

\subsection{Infalling star cluster: eccentric initial orbits}\label{sec:infall}

\cite{Levin05} have recently modelled the dynamics of the stars peeled
off from the IMBH inspiralling to smaller radii in the field of
\sgra. The orbits of these stars were found to gain significant
eccentricities due to repetitive interactions with the IMBH while they
have not yet distanced themselves very far from it.  The effect is the
strongest if the IMBH is itself on an eccentric orbit, when some of
the stars are flung to orbits with an angular momentum opposing their
initial one.  Some of the stars in fact become unbound and escape to
infinity. The eccentric star cluster inspiral thus defies the purpose
of the model -- to bring massive young stars in -- as many are flung
out on orbits in which they spend most of the time outside the central
parsec. This would contradict observations \citep{Genzel03a}. We thus
consider only the circular cluster inspiral model here.  The stars
then obtain\citep{Levin05} relatively mild eccentricities, with a
median of around 0.5, and the resulting stellar disc is rather thin
since the stellar velocities in the direction perpendicular to the
IMBH inspiral orbit are quite small.

\begin{figure}
\centerline{\psfig{file=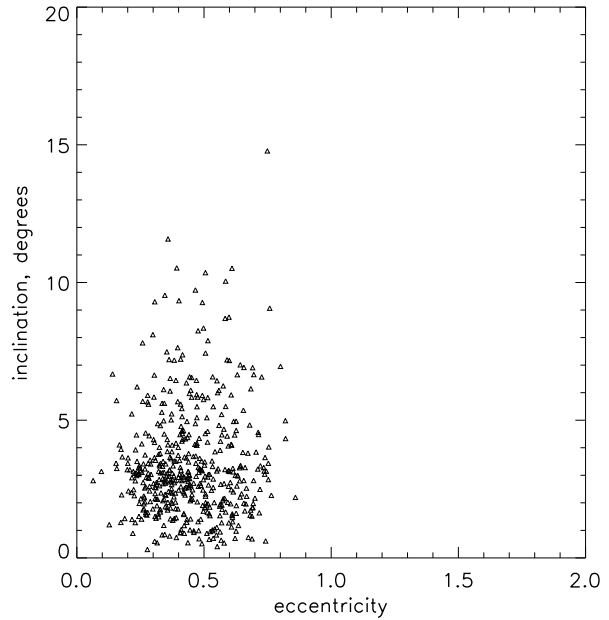,width=.5\textwidth,angle=0}}
\caption{The initial distribution of eccentricities and inclination
angles for a test with 500 particles in the clock-wise (disc)
system. The distribution is chosen to mimic that resulting from a
circular inspiral of an IMBH-star cluster \citep[see Fig. 1 in][]{Levin05}.}
\label{fig:levindisc}
\end{figure}

To generate initial orbits similar to those obtained by 
\cite{Levin05}, we place the stars radially in the same way as before,
but we now add to their velocities random components both in the plane of the disc and perpendicular to the disc. In the plane of the disc, we added a random velocity component ( in both radial and azimuthal directions), $\Delta {\bf v}$,  distributed between $-(1/5) v_c$ to $+(1/5) v_c$, where $v_c$ is the local
circular speed.  A smaller random component in the direction perpendicular to the discs was also given to imitate distribution of orbital inclinations to the system's plane. The resulting distribution of stellar initial eccentricities and inclinations (here defined with respect to their respective disc planes), shown in Figure \ref{fig:levindisc}, is indeed similar to  Figure 1 of \cite{Levin05}.

\begin{table}
\caption{Simulations results for eccentric initial orbits with larger ring.}
\begin{tabular}{@{}lllrrr@{}}
\hline
$M_{\rm disc}$ & $M_{\rm ring}$ & run & $\chi^2_{\rm disc}$ & $\chi^2_{\rm ring}$ & fit quality \\
\hline
3500 & 1750 & EL1 & 1.6 & 3.8 & +\\  
6300 & 1750 & EL2 & 1.5 & 9.5 & + \\  
12000 & 1750 & EL3 & 2.9 & 6.9 & - \\  
3500 & 3500 & EL4 & 2.2 & 1.7 & + \\ 
6300 & 3500 & EL5 & 1.9 & 7.2 & - \\ 
12000 & 3500 & EL6 & 2.7 & 13.0 & - \\ 
35000 & 3500 & EL7 & 15.6 & 3.4 & - \\ 
3500 & 10500 & EL8 & 7.9 & 2.8 & - \\ 
3500 & 14000 & EL9 & 17.5 & 2.7 & - \\ 
12000 & 10500 & EL10 & 8.5 & 12.2 & - \\ 
6650 & 10500 & EL11 & 5.1 & 5.5 & - \\ 
3500 & 7000 & EL12 & 3.0 & 2.1 & - \\ 
\hline
\label{tab:levin}
\end{tabular}
\end{table}

The results of the tests with the given eccentricity and inclination distribution (Figure \ref{fig:levindisc}) 
are summarised in Table \ref{tab:levin}.  The resulting $\chi^2$ values are significantly
larger than those for the initially circular orbits at same masses. We
estimate that the disc masses must be less than
\begin{eqnarray}
M_{\rm disc} \simlt 5 \times 10^3 \msun\\
M_{\rm ring} \simlt 5 \times 10^3 \msun \;.
\label{circ}
\end{eqnarray}
We interpret these tighter upper limits as a result of a wider spectrum of orbits in the eccentric discs. The greater diversity in the orbits leads to a greater difference in the rates at which the orbital planes precess.

\subsection{The maximum current mass of the stellar
systems}\label{sec:now} 

Using the same ideas described above, we can ask a slightly different
question. We can start with orbits spread around in the velocity space
sufficiently wide to yield $\chi^2$ values consistent with the
observed discs, follow these orbits for a shorter time, say a million
year, and then measure the $\chi^2$ again. If $\chi^2$ significantly
increases during this ``short'' time, such a disc should be
rejected. The argument here is that during a time considerably shorter
than the age of the discs, the observed system should not evolve (get
warped and mixed) too much. We have ran a grid of models corresponding
to disc and ring radii from the "larger" ring tests. The approximate
upper limits on the masses of the system obtained in this way are
\begin{eqnarray}
M_{\rm disc} \simlt 1 \times 10^4 \msun\\
M_{\rm ring} \simlt 5 \times 10^3 \msun \;.
\label{mnow}
\end{eqnarray}

\section{Discussion and Conclusions}\label{sec:conclusions}

In this paper we modelled N-body evolution of two stellar discs guessing their
initial geometrical arrangement based on their present day observed
configuration, and then following stellar orbits for 3 Million years. Given
the uncertainty in the initial guess, these tests cannot be precise, but since
the discs are about twice older than 3 Million years \citep{Paumard05}, we
feel our upper mass limits are conservative. In summary, we found that the
total mass of each of the stellar systems orbiting \sgra\ cannot be greater
than $M\simlt (5 - 15) \times 10^3 \msun$, with the lower values for eccentric
initial stellar orbits, and the higher limits for the circular orbits. It is
somewhat disappointing to us that these limits are consistent with both of the
models for star formation near \sgra. The minimum gaseous mass of the disc at
which it becomes gravitationally unstable and forms stars is around $M_{\rm
disc} \simeq 10^4 \msun$ for \sgra\ case \citep{NC05}. Note that this was
derived from the basic \cite{Shakura73} model, without including self-gravity
into the hydrostatic balance equation for the disc. We expect the minimum
$M_{\rm disc}$ to be another factor of $\sim 2$ lower, therefore. The mass of
the stars formed through the disc collapse should be close to the original gas
mass, as argued by \cite{NC05}, because the viscous time scale for gas
accretion is much longer than the time scale on which the stars can devour the
disc. Thus the minimum stellar mass in that model is of order $5 \times 10^3
\msun$, in line with the upper limits obtained here.

Now, for the cluster infall model, the initial mass of the cluster should be
as large as $10^5-10^6\msun$ to allow it to spiral in rapidly enough, as well
as to form an intermediate mass black hole that is heavy enough to prevent
cluster dissolution away from the central parsec
\citep{Kim04,Gurkan05}. However the final stellar masses that make it into the
central parsec are always a very small fraction of the original cluster mass,
and they appear to be consistent with the limits obtained here.

We may also approach these mass limits from another angle. Estimating the
total mass of the {\em observed} young massive stars, and assuming a standard
\cite{Salpeter55} IMF, one would predict the total stellar mass to be around
$\sim$~(few to 10)~$\times 10^4\msun$, depending on the low mass cutoff in the
power-law. This would be a factor of several larger than the mass limits for
the circular initial orbits. Thus, {\em if} the stars were formed inside a
massive self-gravitating disc, their IMF should be at least somewhat deficient
in the low mass stars.  In fact, in a separate paper, \cite{NS05} show that a
similar but even stronger argument can be made on purely observational
grounds.

We thank an anonymous referee for his useful comments.  This research was
supported in part by the National Science Foundation under Grant
No. PHY99-07949 during SN's visit to the Kavli Institute for Theoretical
Physics in Santa Barbara.

\end{document}